\documentstyle[11pt,fleqn]{article}

\newcommand{\be}{\begin{eqnarray}}
\newcommand{\ee}{\end{eqnarray}}

\def\ap[#1,#2]{a_#1 + a_#2}
\def\y{\lambda}

\begin{document}
\title{Solutions of the reflection equation for face and vertex models
associated with $A_n^{(1)},B_n^{(1)},C_n^{(1)},D_n^{(1)}$ and
$A_n^{(2)}$} 
\author{M. T. Batchelor$^{\rm a}$, V. Fridkin$^{\rm a}$, 
A. Kuniba$^{\rm b}$ and Y. K. Zhou$^{\rm a}$\\\\
$^{\rm a}$ Department of Mathematics, School of Mathematical Sciences,\\ 
Australian National University, Canberra ACT 0200, Australia\\
$^{\rm b}$ Institute of Physics, University of Tokyo, Komaba, Meguro-ku,\\ 
Tokyo 153 Japan}

\maketitle
\pagenumbering{arabic}
\begin{abstract}
We present new diagonal solutions of the reflection equation
for elliptic solutions of the star-triangle relation.  The models
considered are related to the affine Lie algebras 
$A_n^{(1)},B_n^{(1)},C_n^{(1)},D_n^{(1)}$ and $A_n^{(2)}$. 
We recover all known diagonal solutions associated with these algebras
and find how these solutions are related in the elliptic regime.
Furthermore, new solutions of the reflection equation follow
for the associated vertex models in the trigonometric limit.
\end{abstract}
\vskip 5 mm
\section{Introduction}
Much work has recently been done in integrable quantum field 
theory \cite{c,gz94b,fk94,jkkkw95,lmss95} and lattice statistical mechanics 
\cite{s88,ku95,z95t,b95t} 
on models with a boundary, where integrability manifests itself via
solutions of the reflection equation (RE) \cite{c,s88}.
In field theory, attention is focused on the boundary $S$-matrix. In  
statistical mechanics, the emphasis has been on  
deriving solutions of the RE and
the calculation of various surface critical phenomena,
both at and away from criticality (see, e.g. \cite{z95t,b95t} for
recent reviews). 

Integrable models exhibit a natural connection with affine Lie 
algebras \cite{j86,b85,jmo87,jmo88,k91}. Our interest here lies in both 
the vertex and face formulation of models associated with the algebras 
$X_n^{(1)}=A_n^{(1)},B_n^{(1)},C_n^{(1)},D_n^{(1)}$ and $A_n^{(2)}$.
These higher rank models include some well known models as
special cases.
In particular, the Andrews-Baxter-Forrester model (ABF) \cite{abf84} 
is related to $A_1^{(1)}$ while the dilute $A_L$ models \cite{wns93}, 
which contain an Ising model in a magnetic field for $A_3$ \cite{wns93},
are related to $A_2^{(2)}$.  
The excess surface critical 
exponents $\alpha_s$ and $\delta_s$ were recently derived for the 
dilute $A_L$ models, including the Ising values 
$\alpha_s=1$ at $L=2$ and $\delta_s=-\frac{15}{7}$ at $L=3$ \cite{bfz95}.  

Here we consider the RE for the higher rank models with a view  
to deriving new surface critical phenomena. The RE was first written down 
in field theoretic language \cite{c}, later for vertex  
models \cite{s88} and more recently in the interaction-round-a-face (IRF) 
formulation \cite{bpo95,fhs95,z95,ku95,ak95}. We have solved the RE for the
above elliptic face models, and in so doing have also obtained new
solutions in the vertex limit. We find an explicit connection 
between solutions found in various limits which were thought isolated before.

\section{IRF formulation and vertex correspondence}

We begin by fixing our notation and recalling the basic ingredients for
integrability, both in the bulk and at a boundary. We will refer 
to solutions of the star-triangle equation (STR) as bulk weights. Given the
bulk weights, we refer to the associated solutions of the RE as
boundary weights.  Our representation of the bulk and boundary weights is 
%
\def\wpic
{\begin{picture}(18,22)(0,0)
\thinlines
\put(2,2){\framebox(18,18){}}
\put(2,10){\line(1,1){10}}
\put(11,10){\makebox(0,0){$u$}}
\end{picture}}
%
%
\def\klpic
{\begin{picture}(18,32)(0,0)
\thinlines
\put(0,0){\line(0,1){32}}
\put(0,32){\line(1,-1){16}}
\put(16,16){\line(-1,-1){16}}
\end{picture}}
\def\kpic
{\begin{picture}(18,32)(0,0)
\thinlines
\put(16,0){\line(0,1){32}}
\put(16,32){\line(-1,-1){16}}
\put(0,16){\line(1,-1){16}}
\put(5,11){\line(0,1){10}}
\put(11,16){\makebox(0,0){$u$}}
 \end{picture}}
%
%
\def\Wp[#1,#2,#3,#4] { \begin{array}{@{}c} 
			  #4 \\ #1 \raisebox{-1.8ex}{ \wpic }\;\; #3 \\ #2  
			\end{array}}
\def\W[#1,#2,#3,#4,#5]{ W \left( \begin{array}{@{}cc|} 
			#1 & #4 \\ #2 & #3 \end{array} \; #5 \right)}
\def\Kp[#1,#2] { \begin{array}{@{}c}
			   #2 \\[-1ex] \kpic  \\[-1.5ex] #1 \end{array}}
\def\K[#1,#2,#3,#4] { K \left( \begin{array}{@{}cc|}
			 & #3 \\ \raisebox{1.2ex}{$#1$}  & #2 \end{array} \; #4 \right)}
%
\be 
\Wp[\mu,\nu,\sigma,\kappa] &=&
\W[a, a+\hat{\mu}, a+\hat{\mu}+\hat{\nu}, a+\hat{\kappa}, u],\qquad
(\hat{\mu}+\hat{\nu} = \hat{\kappa}+\hat{\sigma}) \\ 
\Kp[\mu,\nu] &=& \K[a,a+\hat{\mu},a+\hat{\nu},u] .   
\ee
We use the same algebraic notation as in \cite{jmo88,k91}, with 
Latin letters ($a,b,\cdots$) for states, Greek letters ($\mu,\nu,\cdots$) 
for elementary vectors and $u$ as spectral parameter.  
The corner triangle locates the state associated with the square
(bulk) and triangular (boundary) face (assumed to be $a$ unless
otherwise stated).  In the arrow representation \cite{z95} it
is the point from which the arrows emanate.  This is made clear in the
following relations, which also show the vertex-face correspondence in the
critical (trigonometric) limit,  
%
\def\arrowsquare
{\begin{picture}(20,23)(0,-3)
\put(0,20){\vector(0,-1){13}}
\put(0,20){\vector(1,0){13}}
\put(0,20){\line(0,-1){20}}
\put(0,20){\line(1,0){20}}
\put(20,20){\vector(0,-1){13}}
\put(0,0){\vector(1,0){13}}
\put(20,20){\line(0,-1){20}}
\put(0,0){\line(1,0){20}}
\put(10,10){\makebox(0,0){$u$}}
\end{picture}}
\def\Surround[#1,#2,#3,#4,#5] { \begin{array}{@{}c} 
			  #4 \\ #1  #5 \;\; #3 \\ #2  
			\end{array}}
\def\vertex
{\begin{picture}(30,32)(0,-2)
\put(0,15){\line(1,0){30}}
\put(15,0){\line(0,1){30}}
\put(15,15){\oval(10,10)[bl]}
\put(7,7){\makebox(0,0){$u$}}
\end{picture}}
%
%
\def\arrowtriangle
{\begin{picture}(18,32)(0,0)
\thinlines
\put(16,0){\line(0,1){32}}
\put(16,32){\line(-1,-1){16}}
\put(0,16){\vector(1,1){10}}
\put(0,16){\vector(1,-1){10}}
\put(0,16){\line(1,-1){16}}
\put(11,16){\makebox(0,0){$u$}}
 \end{picture}}
\def\halfvertex
{\begin{picture}(18,32)(0,0)
\thinlines
\put(18,16){\line(-1,1){14}}
\put(18,16){\line(-1,-1){14}}
\put(13,4){\makebox(0,0){$u$}}
\put(18,16){\line(0,-1){10}}
\put(18,12){\oval(7,5)[bl]}
 \end{picture}}
\def\Semisurround[#1,#2,#3]
{ \begin{array}{@{}c}  #2 \\[-1ex] #3 \\[-1.5ex] #1 \end{array}}
\be
\begin{array}{c} 
\Surround[\alpha,\beta,\gamma,\delta,\raisebox{-1.8ex}{ \arrowsquare
  }]  
\; = \;\;   
\Wp[\alpha,\beta,\gamma,\delta] 
\; \stackrel{``|a| \rightarrow \infty''}
{\Longrightarrow}\; 
\Surround[\alpha,\beta,\gamma,\delta,\raisebox{-2.7ex}{ \vertex }]
\; = \; 
R_{\;\alpha\beta}^{\;\delta\gamma}(u),
\\[5.0ex] \quad
\Semisurround[\alpha,\beta, \arrowtriangle ]
\; = \;\;\;
\Kp[\alpha,\beta] 
\;\; \stackrel{``|a| \rightarrow \infty''}
{\Longrightarrow}\!\! 
\Semisurround[\alpha,\beta, $\hskip 5 mm$ \halfvertex ]
\; = \; 
K_{\;\alpha}^{\;\beta}(u) . 
   \end{array}
\label{fv}
\ee

Here the limit ``$|a| \rightarrow \infty$'' will be specified
more concretely in Sec. 5.
The STR has the graphical form


\def\hex
{\begin{picture}(70,40)(0,0)
\put(0,20){\line(1,1){20}}
\put(0,20){\line(1,-1){20}}
\put(70,20){\line(-1,1){20}}
\put(70,20){\line(-1,-1){20}}
\put(20,0){\line(1,0){30}}
\put(20,40){\line(1,0){30}}
\end{picture}}

\def\trileft
{\begin{picture}(70,40)(0,0)
\put(0,20){\line(1,1){20}}
\put(0,20){\line(1,-1){20}}
\put(70,20){\line(-1,1){20}}
\put(70,20){\line(-1,-1){20}}
\put(20,0){\line(1,0){30}}
\put(20,40){\line(1,0){30}}

\put(30,20){\circle*{4}}

\put(27,40){\line(-2,-1){14}}
\put(15,20){\line(-2,-1){10}}
\put(45,35){\line(1,0){10}}
\put(25,30){\makebox(0,0){$u$}}
\put(25,10){\makebox(0,0){$v$}}
\put(50,20){\makebox(0,0){$u-v$}}
\put(0,20){\line(1,0){30}}
\put(50,0){\line(-1,1){20}}
\put(50,40){\line(-1,-1){20}}
 \end{picture}}

\def\triright
{\begin{picture}(70,40)(0,0)
\put(0,20){\line(1,1){20}}
\put(0,20){\line(1,-1){20}}
\put(70,20){\line(-1,1){20}}
\put(70,20){\line(-1,-1){20}}
\put(20,0){\line(1,0){30}}
\put(20,40){\line(1,0){30}}

\put(40,20){\circle*{4}}

\put(35,40){\line(-2,-1){10}}
\put(47,20){\line(-2,-1){14}}
\put(15,35){\line(1,0){10}}
\put(45,30){\makebox(0,0){$v$}}
\put(45,10){\makebox(0,0){$u$}}
\put(20,20){\makebox(0,0){$u-v$}}
\put(70,20){\line(-1,0){30}}
\put(20,0){\line(1,1){20}}
\put(20,40){\line(1,-1){20}}
 \end{picture}}


\be \hskip 30 mm
\begin{array}{ccc} \trileft & \raisebox{3.2ex}{$=$} & \triright
\end{array}
\ee

\noindent where the edges (sites) of the outer hexagon take on the 
same elementary vectors (states) 
on either side of the relation and the internal edges (sites) are
summed over (represented by a full dot).
Once a configuration of vector differences is specified
on the outer hexagon, only one state ($f$ say at the top left corner) 
is required to
specify the others (as with any configuration of meeting faces).

If we assign states anticlockwise in alphabetical order from the left 
most corner of the above hexagon, with summed height $g$ in the centre 
and state $a$ at the start, then the STR reads
\be \begin{array}{l}
\displaystyle{\sum_g} \W[f,a,g,e,u] \W[a,b,c,g,v] \W[e,g,c,d,u-v] = \\[3.0ex] 
\hphantom{\W[f,a,g,e,u]}
\displaystyle{\sum_g} \W[f,a,b,g,u-v] \W[g,b,c,d,u] \W[f,g,d,e,v].
   \end{array}
\label{str}
\ee
The RE has the graphical form
\def\rretri
{\begin{picture}(60,80)(0,0)
\put(20,40){\line(1,-1){40}}
\put(20,40){\line(1,1){40}}
\put(60,0){\line(0,1){80}}
\end{picture}}
\def\rreboxup
{\begin{picture}(60,80)(0,0)
\put(60,40){\line(-1,-1){20}}
\put(60,40){\line(-1,1){40}}
\put(0,60){\line(1,1){20}}
\put(0,60){\line(1,-1){20}}
 \end{picture}}
\def\rreboxdown
{\begin{picture}(60,80)(0,0)
\put(60,40){\line(-1,1){20}}
\put(60,40){\line(-1,-1){40}}
\put(0,20){\line(1,1){20}}
\put(0,20){\line(1,-1){20}}
 \end{picture}}
\def\rrespecsdown
{\begin{picture}(60,80)(0,0)
\put(20,20){\makebox(0,0){$u-v$}}
\put(40,40){\makebox(0,0){$u+v$}}
\put(53,20){\makebox(0,0){$u$}}
\end{picture}}
\def\rreup
{\begin{picture}(60,80)(0,0)
\put(20,40){\line(1,-1){40}}
\put(20,40){\line(1,1){40}}	
\put(60,0){\line(0,1){80}}
\put(60,40){\line(-1,-1){20}}
\put(60,40){\line(-1,1){40}}
\put(0,60){\line(1,1){20}}	
\put(0,60){\line(1,-1){20}}
\put(20,60){\makebox(0,0){$u-v$}}
\put(40,40){\makebox(0,0){$u+v$}}	
\put(53,20){\makebox(0,0){$v$}}		
\put(53,60){\makebox(0,0){$u$}}
\put(15,45){\line(1,0){10}}
\put(25,35){\line(0,1){10}}
\put(45,15){\line(0,1){10}} 
\put(45,55){\line(0,1){10}}
\put(40,60){\circle*{4}}  
\put(60,40){\circle*{4}}
\multiput(30,75)(5,0){5}{\circle*{1}}
\end{picture}}
\def\rredown
{\begin{picture}(60,80)(0,0)
\put(20,40){\line(1,-1){40}}
\put(20,40){\line(1,1){40}}	
\put(60,0){\line(0,1){80}}
\put(60,40){\line(-1,1){20}}
\put(60,40){\line(-1,-1){40}}
\put(0,20){\line(1,1){20}}	
\put(0,20){\line(1,-1){20}}
\put(20,20){\makebox(0,0){$u-v$}}
\put(40,40){\makebox(0,0){$u+v$}}	
\put(53,20){\makebox(0,0){$u$}}		
\put(53,60){\makebox(0,0){$v$}}
\put(15,35){\line(1,0){10}}
\put(25,35){\line(0,1){10}}
\put(45,15){\line(0,1){10}} 
\put(45,55){\line(0,1){10}}
\put(40,20){\circle*{4}}  
\put(60,40){\circle*{4}}
\multiput(30,5)(5,0){5}{\circle*{1}}	
\end{picture}}
\be \hskip 30 mm 
\begin{array}{ccc} \rreup & \raisebox{6.8ex}{ $=$} & \rredown 
\end{array}
\ee  
\noindent where the external edges of same state on each side of the relation 
carry the same vector differences and internal edges are summed.  The
edges connected by dashes are identified as one internal edge.  

If we assign states as before but starting from the top most corner
and sum in both $f$ and $g$ we have the RE in the form
\be \begin{array}{l} \displaystyle{\sum_{fg}} \,\,
\W[c,g,a,b,u-v] \K[g,f,a,u] \W[c,d,f,g,u+v] \K[d,e,f,v] = \\[3.0ex]
\hphantom{WW} \displaystyle{\sum_{fg}} \,\,
\K[b,f,a,v] \W[c,g,f,b,u+v] \K[g,e,f,u] \W[c,d,e,g,u-v].
   \end{array}
\label{re}
\ee
This formulation follows that given in \cite{fhs95,z95,ku95,ak95}. It
is equivalent to the formulation of \cite{bpo95} in the special case that a
crossing symmetry of the bulk weights exists. Such crossing symmetry is
absent in the higher rank $A^{(1)}_{n>1}$ models \cite{jmo87} 
and as a result the formulation given in \cite{bpo95} does not support 
diagonal boundary weight solutions for these models.    

The above RE (\ref{re}) reduces to the original 
formulation \cite{c,s88} 
\be
R_{12}(u-v)K_1(u)R_{21}(u+v)K_2(v) 
 =K_2(v)R_{21}(u+v)K_1(u)R_{12}(u-v),
\label{rev}
\ee
in the vertex limit (\ref{fv}).

\section{Bulk face weights}

Consider the 
elliptic face models associated with 
$X^{(1)}_n$ \cite{jmo88} and $A^{(2)}_n$ \cite{k91}.
In these models, the states range over the dual space of 
the Cartan subalgebra of $X^{(1)}_n$ and ${\dot X}^{(1)}_n$ \cite{k91}, 
respectively.
Arrows $\alpha, \beta, \mu, \nu$, etc  run over the set
\be 
\begin{array}{ll}
\{ 1, 2, \ldots, n+1 \} &  \mbox{ for  }
A_n^{(1)},
\\[2.0ex] 
\{ 1, 2, \ldots, n, 0, -n, \ldots, -1 \}  & \mbox{ for  }
B_n^{(1)}, A^{(2)}_{2n}, \\[2.0ex]
\{ 1,2, \ldots, n, -n, \ldots, -1 \}  & \mbox{ for  }
C_n^{(1)}, D^{(1)}_n,  A^{(2)}_{2n-1}.
   \end{array}
\label{arrow}
\ee
In particular, $\sum_\kappa, \prod_\beta$, etc are to be 
taken over the above set.
In terms of the notation of \cite{jmo88,k91} the bulk weights are
given by
\def\a[#1,#2]{a_#1 - a_#2}
\def\gap{\hskip 5 mm}
\[ 
\Wp[\mu,\mu,\mu,\mu] = \frac{[1+u]}{[1]}, \gap
\Wp[\mu,\nu,\nu,\mu] = \frac{[\a[\mu,\nu]-u]}{[\a[\mu,\nu]]}\gap (\mu\ne\nu),
\]

\[
\Wp[\mu,\nu,\mu,\nu] = \frac{[u]}{[1]}\left(
		\frac{[\a[\mu,\nu]+1][\a[\mu,\nu]-1]}{[\a[\mu,\nu]]^2}
			\right)^{1/2} \gap (\mu\ne\nu).
\]
for $A^{(1)}_n$, while for $B^{(1)}_n,C^{(1)}_n,D^{(1)}_n, A_n^{(2)}$
they read
\[ 
\Wp[\mu,\mu,\mu,\mu] = \frac{[\y-u][1+u]}{[\y][1]} \gap (\mu\ne 0),
\]

\[
\Wp[\mu,\nu,\nu,\mu] = \frac{[\y-u][\a[\mu,\nu]-u]}{[\y][\a[\mu,\nu]]}\gap
(\mu\ne\pm\nu),
\]

\[
\Wp[\mu,\nu,\mu,\nu] = \frac{[\y-u][u]}{[\y][1]}\left(
		\frac{[\a[\mu,\nu]+1][\a[\mu,\nu]-1]}{[\a[\mu,\nu]]^2}
			\right)^{1/2} \gap (\mu\ne\pm\nu),
\]

\[
\Wp[\mu\;\;,-\mu\;,-\nu,\nu\;] =
\frac{[u][\ap[\mu,\nu]+1+\y-u]}{[\y][\ap[\mu,\nu]+1]} 
		(G_{a,\mu}G_{a,\nu})^{1/2}\gap (\mu\ne\nu), 
\]

\[
\Wp[\mu\;\;,-\mu\;,-\mu,\mu\;] =
	\frac{[\y+u][2a_\mu+1+2\y-u]}{[\y][2a_\mu+1+2\y]} -
	\frac{[u][2a_\mu+1+\y-u]}{[\y][2a_\mu+1+2\y]} H_{a,\mu},
\]

\[
\hphantom{ \Wp[\mu\;\;,-\mu\;,-\!\mu,\mu\;] } =
	\frac{[\y-u][2a_\mu+1-u]}{[\y][2a_\mu+1]} +
 	\frac{[u][2a_\mu+1+\y-u]}{[\y][2a_\mu+1]}G_{a,\mu}(\mu\ne0). 
\]

In the above we remind the reader that
$a_\mu = -a_{-\mu}$ for all $\mu$ in (\ref{arrow}) except
$a_0 = -1/2$ \cite{jmo88,k91}.
The crossing parameter is given by $\lambda = -tg/2$ for $X_n^{(1)}$ and
$\lambda  = -g/2 + L/2$ for $A_n^{(2)}$
(note that $\lambda$ is shifted by $L/2$ from \cite{k91}).
The parameters $t,g$ are given in Table 1.
Here $L$ is arbitrary for the unrestricted solid-on-solid (SOS) models but
will be specified later for the restricted (RSOS) models. We have
further defined
\be
[u] = [u,p] = \vartheta_1(\pi u /L,p),
\ee
where
\be
\vartheta_1(u,p) = 2|p|^{1/8} \sin u {\displaystyle\prod_{j=1}^\infty} (1 - 2p^
j \cos 2u + p^{2j}) (1-p^j)
\ee
is a standard elliptic theta-function of nome $p = e^{2\pi i\tau}$.
For convenience we also use 
\be
[u,p]' = \vartheta_4(\pi u/L,p),
\ee
where
\be
\vartheta_4(u,p) = {\displaystyle\prod_{j=1}^\infty} (1 - 2p^{j-1/2}
\cos 2u + p^{2j-1}) (1-p^j).
\ee
The quantity $G_{a, \mu}$ 
is determined by 
$G_{a, \mu} = G_{a+\hat{\mu}}/G_a\;\;(\mu\ne 0)$  and $ G_{a,0} = 1$,
where
\be \begin{array}{l}
G_a = {\displaystyle\prod_{1\le i<j\le n+1}} [a_i-a_j] \mbox{\ \ \ for
$A_n^{(1)}$,} \\[4.0ex]
\hphantom{G_a } = \varepsilon(a) {\displaystyle\prod_{i=1}^n} h(a_i)
{\displaystyle\prod_{1\le i<j\le n}} [a_i-a_j] [a_i+a_j] 
\mbox{\ \ \ otherwise.}
   \end{array}
\ee
The sign factor $\varepsilon(a)$ is such that
$\varepsilon(a+\hat{\mu})/\varepsilon(a) = -1$ for $C_n^{(1)}$ and
$A_{2n-1}^{(2)}$ only and is unity for the other cases. The function
$h(a)$ is given in Table 1. 
Finally,
\be
H_{a,\mu} = \sum_{\kappa\ne\mu}
        \frac{[\ap[\mu,\kappa]+1+2\y]}{[\ap[\mu,\kappa]+1]}G_{a,\kappa} .
\ee

\vskip 5 mm

\begin{tabular}{@{}*{8}{c}}
{\bf Table 1}\\  \hline \\

type & $A_n^{(1)}$ & $B_n^{(1)}$ & $C_n^{(1)}$ & $D_n^{(1)}$ &
$A_{2n}^{(2)}$ & $A_{2n-1}^{(2)}$ \\[2.0ex]

$g$ & $n+1$  & $2n-1$  & $n+1$
 & $2n-2$  & $2n+1$ & $2n$ \\[2.0ex]

$t$ & 1  & 1 & 2
 & 1 & 1 & 2 \\[2.0ex]

$h(a)$ & 1 & $[a]$ & $[2a]$ & 1 & $[a][2a,p^2]'$ & $[2a,p^2]$\\[2.0ex] \hline
\end{tabular}

\vskip 10 mm

For the face models two inversion 
relations are satisfied by the bulk weights \cite{jmo88,k91},
\def\twosquare
{\begin{picture}(80,40)(0,0)
\put(0,20){\line(1,1){20}}
\put(0,20){\line(1,-1){20}}
\put(20,0){\line(1,1){40}}
\put(20,40){\line(1,-1){40}}
\put(80,20){\line(-1,1){20}}
\put(80,20){\line(-1,-1){20}}
\end{picture}}
\def\inversionA
{\begin{picture}(80,40)(0,0)
\put(0,20){\line(1,1){20}}
\put(0,20){\line(1,-1){20}}
\put(20,0){\line(1,1){40}}      
\put(20,40){\line(1,-1){40}}
\put(80,20){\line(-1,1){20}}
\put(80,20){\line(-1,-1){20}}
\put(15,35){\line(1,0){10}}
\put(55,35){\line(1,0){10}}     
\put(40,20){\circle*{4}}
\put(20,20){\makebox(0,0){$u$}}         
\put(60,20){\makebox(0,0){$-u$}}
\multiput(30,5)(5,0){5}{\circle*{1}}    
\multiput(30,35)(5,0){5}{\circle*{1}}
\end{picture}}
\def\inversionB
{\begin{picture}(80,40)(0,0)
\put(0,20){\line(1,1){20}}
\put(0,20){\line(1,-1){20}}
\put(20,0){\line(1,1){40}}      
\put(20,40){\line(1,-1){40}}
\put(80,20){\line(-1,1){20}}
\put(80,20){\line(-1,-1){20}}
\put(45,15){\line(0,1){10}}
\put(35,15){\line(0,1){10}}     
\put(40,20){\circle*{4}}
\put(20,20){\makebox(0,0){$\lambda-u$}}         
\put(60,20){\makebox(0,0){$\lambda+u$}}
\multiput(30,5)(5,0){5}{\circle*{1}}    
\multiput(30,35)(5,0){5}{\circle*{1}}
\end{picture}}
\be \displaystyle{\sum_g}\;
 \W[a,b,c,g,u] \W[a,g,c,d,-u] &=& \delta_{bd} \varrho_1(u), \\
 \displaystyle{\sum_g}
\left( {\displaystyle\frac{G_g G_b}{G_a G_c}} \right)
        \W[g,a,b,c,\lambda-u] \W[g,c,d,a,\lambda+u] &=& \delta_{bd}
\varrho_2(u).
\ee
Here the inversion functions are given by
\be
\varrho_1(u) &=& {\displaystyle\frac{[1+u][1-u]}{[1]^2}},\;\;\;
   \varrho_2(u) = {\displaystyle\frac{[\lambda+u][\lambda-u]}{[1]^2}} \quad
        \mbox{for $A_n^{(1)}$}, \\
\varrho_1(u) &=& \varrho_2(u) =
{\displaystyle\frac{[\lambda+u][\lambda-u][1+u][1-u]}{[\lambda]^2[1]^2}}
\qquad \mbox{otherwise.}
\ee

\section{Boundary face weights}

Given the above solutions of the STR (\ref{str}), we have found that the 
RE (\ref{re}) has the diagonal face solutions
\be 
\K[a+\hat{\mu},b,a,u] = 
{\displaystyle\frac{[a_\mu-\eta+u]}
{[a_\mu-\eta-u]}}\; 
 f_a(u)\;\delta_{a b},
\label{sol} 
\ee
For $A_n^{(1)}$, $\eta$ is a free parameter. However, 
for $B^{(1)}_n,C^{(1)}_n,D^{(1)}_n$ and $A_n^{(2)}$,
it must be chosen as 
\be
\eta = -{\lambda + 1 \over 2} + {rL \over 2}
 + {sL\tau \over 2}, \quad r, s \in {\bf Z}.
\label{rs}
\ee
The two solutions (\ref{sol}) corresponding to $(r,s)$ and $(r',s')$ 
are in fact different only by an overall function of $u$ if
$r-r' \equiv s-s' \equiv 0$ mod 2.
The function $f_a(u)$ is not restricted from the 
RE (\ref{re}).
However, it may be normalized as
\be 
f_a(u) = 
z_a(u) e^{-2\pi i\omega s u/L} 
h(u+\eta)\prod_{\beta}[a_\beta-\eta-u],
\label{fdef}
\ee
where $\omega$ is defined by
\be \omega = \left\{ \begin{array}{ll}
	 0  & \quad A_1^{(1)}\\[2.5ex] 
	 \lambda & \quad C_n^{(1)}, D^{(1)}_n\\[2.5ex] 
	 \lambda-1/2 & \quad B_n^{(1)}\\[2.0ex]
         \lambda-1/2-L/2 & \quad A_{2n}^{(2)} \\[2.5ex]
         \lambda-L/2 & \quad A_{2n-1}^{(2)}
			 \end{array} \right. 
\ee
and $z_a(u)$ is any function satisfying 
$z_a(u+\lambda) = z_a(-u)$.
Then the solutions (\ref{sol}) further fulfill
the boundary crossing relation
(except $A_{n>1}^{(1)}$)  
\be
\sum_g \left( {\displaystyle\frac{G_g}{G_b}} \right)^{1/2}
\W[a,b,c,g,2u+\lambda] \K[g,c,a,u+\lambda] = \varrho_3(u) \K[b,c,a,-u] .
\label{bcr}
\ee
Here the boundary crossing function $\varrho_3(u)$ is given by
\be
 \varrho_3(u) =
       \left\{ \begin{array}{ll}
        {\displaystyle\frac{[2-2u]}{[1]}}& \mbox{for\ } A_1^{(1)},\\
        {\displaystyle\frac{[2u+2\lambda][1-\lambda-2u]}{[1][\lambda]}} &
        \mbox{for\ } B_{n}^{(1)}, C_{n}^{(1)}, D_{n}^{(1)} \mbox{and\ }
                     A_{n}^{(2)}.
                         \end{array} \right.
\label{rhos} 
\ee
The proof of (\ref{re}) and (\ref{bcr}) for 
({\ref{sol}) is similar to that for the STR in
\cite{jmo88}.
In the sequel we shall exclusively consider the cases
$r, s = 0, 1$ without loss of generality.

\section{From face to vertex weights}

In the limit $p \rightarrow 0$,
$\vert a_\mu \vert \rightarrow \infty$ ($\mu \neq 0$)
the bulk weights in Sec. 3 reduce, up to normalization,
to the vertex Boltzmann weights given in appendix A of \cite{k91}
for $A^{(2)}_{2n-1}$ and in \cite{j86,b85} for the other algebras.
In the correspondence (\ref{fv}), the indices (\ref{arrow}) 
should be identified, including their orders, with 
$1,2, \ldots, N$ in \cite{j86}.
The parameters $k=e^{-2h}, x, \xi$ in \cite{j86} and
$L, u, \lambda$ here are related as
$h = i \pi/2L, k=e^{-i\pi/L},
x=e^{2\pi iu/L}, \xi=e^{2\pi i \lambda/L}$.
In terms of these variables,
the limit is to be taken so that
$|k^{a_i}| \ll |k^{a_j}|, (i<j)$ for $A^{(1)}_n$ and 
$|k^{a_1}| \ll |k^{a_2}| \ll \cdots \ll |k^{a_n}| \ll 1$
for the other algebras.
Proceeding in the same manner for the boundary 
weights, we obtain new diagonal $K$-matrices that
satisfy the RE (\ref{rev}) 
and the trigonometric limits of (\ref{bcr}) and (\ref{rhos}).  
Below we shall present them using the function
$\epsilon_z = -1+2|\mbox{sgn}(z)|$ with
$\mbox{sgn}(z) = 1,0$ and $-1$ for $z>0,z=0$ and $z<0$, respectively.
\par
The RE for $A_n^{(1)}$ has the diagonal vertex solutions
\be
K_{\;\alpha}^{\;\alpha}(u) = F(u) 
     e^{4 \mbox{sgn}(\alpha - \kappa) h u} 
     \sinh [2h(\phi + \epsilon_{\alpha - \kappa}u)],
\label{av}
\ee
where $\phi \in {\bf C}$, $\kappa \in {\bf R}$ and $F(u)$ are
arbitrary.
Note that for
non-integer $\kappa$ there are essentially only exponential solutions.  
Since any RE in the vertex limit 
involves only two boundary weights and the sign of their state
difference, the following (cf. $\kappa = 1$ or $n+1$) solution is also
admissible
\be 
K_{\;\alpha}^{\;\alpha}(u) = 
\left\{ \begin{array}{ll}
F(u) e^{-2hu} \sinh[2h(\phi+u)] & \alpha \le \kappa, 
\\[2.0ex] 
F(u) e^{2hu} \sinh[2h(\phi-u)] & \alpha > \kappa.
\end{array} \right.
\label{vega}
\ee
This recovers the solution for the $A_n^{(1)}$   
vertex model given in \cite{vg93b} if we take $h=1/2$.   
The function $F(u)$ 
is not restricted by the RE. 
However for $A^{(1)}_1$ ($\lambda = -1$), its appropriate choice
renders the vertex analogue of (\ref{bcr}) still valid 
with the same $\varrho_3(u)$ and $[u] \propto\sinh(2hu)$.
Up to an overall factor $z(u)$ obeying $z(u-1)=z(-u)$,
there are three such normalized solutions.
One of them is given by putting $\kappa = 1$ and $F(u) = 1$ in (\ref{vega}).
The other two are
$(K^1_1(u), K^2_2(u)) = (1,1)$ and $(e^{-4hu},e^{4hu})$.
\par
For the remaining algebras we classify the 
vertex limit according to whether the integer $s$ in (\ref{rs}) is 0 or 1.  
\vskip 5mm
\noindent
{\boldmath{$s$}} {\bf = 0}.
\be 
K^\alpha_\alpha(u) = \left\{ \begin{array}{ll}
F(u) e^{-4\mbox{sgn}(\alpha) h  u}
\sinh[2h({\lambda \over 2}-\epsilon_\alpha u)+{i\pi r\over 2}]
     \sinh[2h({\lambda +1 \over 2}-u)+{i\pi r \over 2}]
  		&\, B_n^{(1)}, A^{(2)}_{2n}\\[2.5ex] 
F(u) e^{-4 \mbox{sgn}(\alpha) h u}\sinh[4h({\lambda + 1 \over 2}-u)]
		&\, C_n^{(1)}, A^{(2)}_{2n-1}\\[2.5ex] 
F(u) e^{-4 \mbox{sgn}(\alpha) h u} &\,D_n^{(1)}
			 \end{array} \right. 
\label{szero}
\ee
\vskip 5mm
\noindent
{\boldmath{$s$}} {\bf = 1}.
In this case we have only $K$ matrices which are
multiples of the identity:
\be 
K^\alpha_\alpha(u) = \left\{ \begin{array}{ll}
F(u) &\, B_n^{(1)}, D^{(1)}_n, A^{(2)}_{2n-1}\\[2.5ex] 
F(u)\sinh[4h({\lambda + 1 \over 2}-u)]
		&\, C_n^{(1)}, A^{(2)}_{2n}
			 \end{array} \right. 
\label{sone}
\ee
As in (\ref{av}) and (\ref{vega}),
$F(u)$ is not restricted from the RE
in (\ref{szero}) and (\ref{sone}).
However,
the simple choice $F(u)=1$
(up to a factor $z(u)$ satisfying $z(u+\lambda) = z(-u)$) allows
the vertex analogue of (\ref{bcr}) to hold with again 
the same $\varrho_3(u)$ and $[u] \propto \sinh(2hu)$.
\par
All the solutions in \cite{mn91a} can be recovered
from our solution for $A_2^{(2)}$.
Besides the identity solution ($s=1$), (\ref{szero}) 
in this case is proportional to
\be 
K_{-1}^{-1}(u) = e^{4hu} \;,\;\;\;\;\;\;
K_0^0 (u) = 
{{\sinh[2h({\lambda \over 2}+u)+{\pi i r \over 2}]} \over 
     {\sinh[2h({\lambda \over 2}-u)+{\pi i r \over 2}]}}
\mbox{\ \ \ and \ \ \ }
K_1^1 (u) = e^{-4hu}\; .
\ee
This can be identified with the two nontrivial solutions in \cite{mn91a}.

\section{Restricted models}

The RSOS models follow in a natural way from the unrestricted models
that we have discussed so far.
To do this, one sets $L = t(l+g)$ for $X^{(1)}_n$ 
($L=t(l+{\dot g})$ for $X^{(2)}_n$) where $t,g$ are given in Table
1 and $l$ is a positive integer.  The local state $a$ is taken as a level 
$l$ dominant integral weight of $X^{(1)}_n$ 
(${\dot X}^{(1)}_n$ for $X^{(2)}_n$).
One also imposes special adjacency rules on the states 
(see \cite{jmo88,k91} for details).
Then it has been shown that the
restricted bulk weights are finite and satisfy the STR 
(\ref{str}) and the inversion relations among 
themselves.  
We have proved that all such features are also valid in 
the boundary case.
Namely, under the restriction 
the boundary face weights (\ref{sol}), (\ref{fdef}) 
(with the simple choice $z_a(u)=1$ for example) are finite
and satisfies 
the RE 
(\ref{re}) and boundary 
crossing relation (\ref{bcr}) among themselves.
The only previous boundary solutions for such RSOS models were for the 
ABF model \cite{bpo95,ak95} 
and for the dilute $A_L$ models built on $A_2^{(2)}$ \cite{bfz95}. 

\section{Summary and conclusion}

We have derived elliptic diagonal solutions of the RE  
for the face models related to the affine Lie algebras 
$A_n^{(1)},B_n^{(1)},C_n^{(1)},D_n^{(1)}$ and $A_n^{(2)}$.  
For $A_n^{(1)}$
we found a free parameter $\eta$ which also survives
in the vertex limit in agreement with previous work \cite{vg93b}.  
For the other algebras we found that $\eta$ takes only discrete values.  
This led to four inequivalent forms of solution corresponding to
$r,s \in {\bf Z}_2$.
Of these four, the two solutions 
for $(r,s) = (0,1)$ and $(1,1)$ degenerated to the identity upon taking 
the vertex limit.\footnote{The solution $K=1$ had previously been known 
for the higher rank cases and leads to quantum group invariant spin 
chains \cite{mn91a,mn91b}.}
This explains the presence of only three distinct solutions in
previous work \cite{mn91a} and shows their connection lies in the 
elliptic regime.
An analogous set of three solutions exist for $G_2^{(1)}$ \cite{yb95b}.  
%
%
For the face models,  previously known solutions were for $A_1^{(1)}$ 
\cite{bpo95,ak95} (see also {\cite{zb95b} 
for related work) and $A_2^{(2)}$ \cite{bfz95}.  These are easily
recovered from those given here.
The transfer matrix eigenspectra for the 
diagonal $K$-matrices can be obtained in the vertex limit 
by means of the Bethe Ansatz, as has been done already for 
a number of models 
(see, e.g. \cite{s88,mn92,vg94,yb95a,yb95b,yb95c,amn95} 
and refs therein).
It is known that different $K$-matrices lead to different
universality classes of surface critical behaviour \cite{by95,yb95c}. 
The investigation of the higher rank solutions obtained here 
is thus of considerable interest.
The elliptic solutions for the higher rank face models are also of 
interest from
the point of view of calculating physical quantities, such as 
surface critical exponents, away from criticality.     

One of the authors (AK) thanks the mathematical physics group at the
Australian National University for their hospitality during his visit.
VF is supported by an Australian Postgraduate Research Award and 
MTB and YKZ are supported by the Australian Research Council.


\end{document}